\documentclass[superscriptaddress,twocolumn,10pt,nofootinbib,aps,floatfix]{revtex4-2}
\usepackage{amssymb,amsmath,times,graphicx,subeqnarray,bm,float}

\usepackage{color}
\usepackage[colorlinks,
linkcolor=blue,
urlcolor=blue,
anchorcolor=blue,
citecolor=blue
]{hyperref}

\begin{document}

\title{Small exciton effective mass in QL Bi$_2$Se$_2$Te: A material platform towards high-temperature excitonic condensate} 

\author{Yuanyuan Wang}
\affiliation
{School of Physics, State Key Laboratory of Crystal Materials, Shandong University, Jinan 250100, China}
\author{Ying Dai}
\affiliation
{School of Physics, State Key Laboratory of Crystal Materials, Shandong University, Jinan 250100, China}
\author{Baibiao Huang}
\affiliation
{School of Physics, State Key Laboratory of Crystal Materials, Shandong University, Jinan 250100, China}
\author{Yee Sin Ang}
\email{yeesin$_$ang@sutd.edu.sg}
\affiliation
{Science, Mathematics and Technology (SMT), Singapore University of Technology and Design, Singapore 487372}
\author{Wei Wei}
\email{weiw@sdu.edu.cn}
\affiliation
{School of Physics, State Key Laboratory of Crystal Materials, Shandong University, Jinan 250100, China}

\begin{abstract}
Using first-principles simulations combined with many-body calculations, we show that two-dimensional free-standing quintuple-layer (QL) Bi$_{2}$Se$_{2}$Te is an inversion symmetric monolayer expected to achieve spatially indirect exciton with large exciton radius, small exciton effective mass and long exciton lifetime. Such system is theoretically predicted to be a promising platform for realizing excitonic Bose–Einstein condensation and superfluid due to its high phase transition temperatures of $\sim$257 K and $\sim$64.25 K for the BEC and excitonic superfluid, respectively. The importance of spin$–$orbit coupling is revealed, and the angular momentum selection rules for photon absorption are discussed. This finding suggests the potential of QL Bi$_{2}$Se$_{2}$Te monolayer with exotic bosonic bound states provides as a tantalizing high-temperature platform to probe excitonic physics.
\end{abstract}

\maketitle

\section{Introduction}
\vspace{-2mm}

Coherent states of excitons have long been a subject of condensed-matter physics. As one of the macroscopic manifestations of quantum coherence, Bose$–$Einstein condensation (BEC) was produced in a vapor of rubidium atoms near a temperature of 170 nK \cite{Anderson269}. Subsequently, exciton polaritons in semiconductor microcavities were verified to undergo a BEC-like phase transition into a superfluid state, a quantum fluid flowing without friction that has a subtle link with BEC \cite{amo2009superfluidity}. In excitonic systems, two intrinsic problems are hindering the realization of BEC: (1) relatively short exciton lifetimes, and (2) dissociation of excitons into unbound electrons and holes. These problems might be circumvented when the Coulomb attraction bounding electrons and holes is sufficiently strong, and, at the same time, the wave functions of electrons and holes have very small overlap. Therefore, coupled quantum wells and van der Waals (vdW) heterostructures of two-dimensional (2D) materials are attracting attention in this field \cite{jiang2021interlayer,fogler2014high, gupta2020heterobilayers, zhu2019gate, conti2021electron, high2012spontaneous}. As a result, spatially indirect excitons with electrons and holes confined into opposite atomic layers are highly sought alternatives for achieving quantum condensation and superfluidity in solid-state devices at experimentally accessible temperatures \cite{fogler2014high, gupta2020heterobilayers, zhu2019gate, conti2021electron, high2012spontaneous, wang2019evidence}. In contrast to atomic systems, the critical temperature of the excitonic BEC in crystal materials can be expected to reach as high as 100 K \cite{fogler2014high, wang2019evidence}.

In quantum wells, however, electron$–$hole pairs are separated usually by dozens of angstroms, and, therefore, the electron–hole Coulomb attraction are rather weak \cite{conti2021electron}. Moreover, surface roughness/disorder also degrades the coherence of excitonic states \cite{gupta2020heterobilayers}. Thus, vdW hetero/homostructures constituted by monolayer materials (\textit{e.g}., transition-metal dichalcogenides, TMDCs) were regard as promising platform to carry excitonic condensation due to the reduced screening and atomically sharp interfaces. In a seminal experiment of electrically generated exciton, the excitonic condensate of electrically generated interlayer excitons were studied in MoSe$_2$–WSe$_2$ atomic double layers. However, the inevitable charged excitons (trions) may have hindered the realization of high-temperature BEC. Therefore, 100 K as the highest critical temperature reported so far \cite{fogler2014high,zhu2019gate,wang2019evidence}. Optically neutral excitons systems may be more appealing than electrical counterpart. However, in the MoSe$_2$–WSe$_2$ photogenerated exciton system, absolute occupation is reported to survive at 10 K and the critical condensation temperature is lower than 100K \cite{Sigl042044}. The lower critical condensation temperature, on the one hand, is attributed to the fact that  the optical interlayer exciton density in TMDCs vdW heterobilayers is rather limited due to the rapid recombination of the intralayer excitons and the long transfer channel \cite{jiang2021interlayer, rivera2018interlayer, chen2016ultrafast}. 
On the other hand, the indirect transitions (considering SOC) caused by strong interlayer interactions induce momentum-indirect low-energy excitons. Usually, strong interlayer interaction and Rashba effect due to the mirror asymmetry result in direct$–$indirect band gap transition of the heterostructure \cite{Sigl042044,niu2021direct}. 
However these momentum-indirect low-energy excitons go against the excitonic condensation \cite{snoke2002spontaneous,Sigl042044,ulman2021organic}. In addition, graphene double bilayers have also been studied in recent theory and experiments. 
Nevertheless, the excitonic condensation is challenging to be observed due to the low critical temperature \cite{perali2013high,li2017excitonic,Berman035418}.

Utilizing a single 2D materials, rather than stacking together forming the vdW heterostructures, to harbor the BEC and superfluidity based on optically generated excitons has yet to be reported thus far. To this end, we propose the following criteria: (1) the electrons and holes constituting the lowest-energy excitonic state should be spatially separated into different atomic layers; (2) the transition dipole moment between the highest valence band and the lowest conduction band is nonzero; (3) direct band gap with small carrier effective mass; (4) good ease of experimental access. In full consideration of these requirements, we screen through multiple 2D candidates in 2D material database \cite{mounet2018two} and theoretically verify that quintuple-layer (QL) Bi$_{2}$Se$_{2}$Te with inversion symmetry is a promising choice to realize BEC condensation and superfluidity truly in two dimensions.

\vspace{-2mm}
\section{Computational Methods}
\vspace{-2mm}

First-principles calculations were carried out with Perdew–Burke–Ernzerhof (PBE) functionals in the framework of generalized gradient approximation (GGA) to study the electronic properties of ground state, by using the Vienna ab initio simulation package (VASP) and Quantum Espresso (QE) \cite{kohn1965self,perdew1996generalized,kresse1996efficient,giannozzi2009quantum,QE-2017}. To mimic an isolated monolayer, a vacuum space was set to 20 \AA. The cut-off kinetic energy for plane waves was set to 500 eV. Structures were fully relaxed until the force on each atom was less than 0.01 eV/\AA$^{-1}$, and the convergence tolerance for energy was 10$^{-5}$ eV. A Monkhorst–Pack $k$-point mesh of 15$\times$15$\times$1 was used to sample the Brillouin zone for geometry optimization and static electronic structure calculations \cite{monkhorst1976special}.
In order to obtain correct excited-state properties and optical absorption spectra, GW approximation combine with Bethe–Salpeter equation (BSE) were used, based on the many-body Green’s function perturbation theory. In particular, PBE energies were corrected by one-shot G$_{0}$W$_{0}$ approximation to attain the quasiparticle eigenvalues. In conjunction with the QE distribution, the GW+BSE calculations were performed by the YAMBO code \cite{marini2009yambo,giannozzi2009quantum}. In order to describe the spin–orbit coupling (SOC) effects, Troullier–Martins norm-conserving fully relativistic pseudopotential was used for the plane wave functions \cite{troullier1991efficient}. 
The plane wave energy cutoff for ground-state property calculations was set to 80 Ry, and a 27$\times$27$\times$1 $k$-grid was used to obtain the quasiparticle energies. As for the self-energy and the dynamical dielectric screening, 500 bands were employed. In order to avoid the long-range interactions, a Coulomb cutoff of the screened potential was used in both G$_{0}$W$_{0}$ and BSE calculations. In the Bethe–Salpeter kernel, four highest valence bands and four lowest conduction bands were considered. In our convergence tests, we confirm that the number of bands and $k$ points were sufficient, such that the change in quasiparticle band gap and excitonic energy in the BSE spectra were within 50 meV.

\begin{figure}
	\centering
	\includegraphics{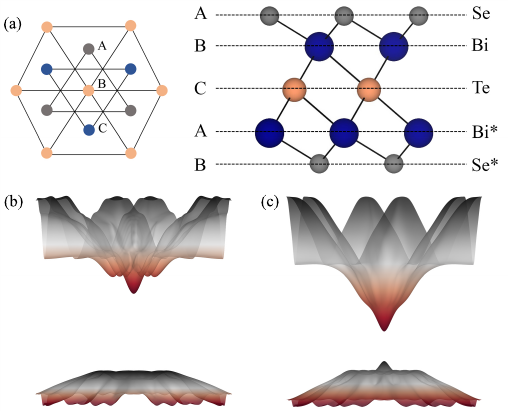}
	\caption{(a) Structure of QL  Bi$_{2}$Se$_{2}$Te from top (left panel) and side (right panel) views. Three different positions in the triangle lattice are denoted as A, B, and C, and atomic layers are labeled. Band structure of QL Bi$_{2}$Se$_{2}$Te in 3D (b) without and (c) with SOC.}
	\label{FIG.1.}
\end{figure}

\vspace{-2mm}
\section{RESULTS AND DISCUSSION}
\subsection{Crystal structure and electronic properties}
\vspace{-2mm}

In experiments and theory, the topological behaviors of layered bulk/film Bi$_{2}$Se$_{2}$Te with a space group of $R\overline{3}m$ ($no.166$) have been widely studied \cite{chang2011density, neupane2012topological,miyamoto2012topological}. As the smallest unit of bulk Bi$_2$Se$_2$Te, QL Bi$_2$Se$_2$Te is a quintuple-layer structure composed of Se-Bi-Te-Bi-Se atomic configuration across its thickness, see Fig.~\textcolor{blue}{1(a)}. Subsequently, we confirm the stability of QL Bi$_{2}$Se$_{2}$Te in terms of exfoliation energy, \textit{ab initio} molecular dynamics (AIMD) simulations and phonon band dispersion (see Appendix \textcolor{blue}{A}). 
The band structure of QL Bi$_{2}$Se$_{2}$Te illustrates an indirect band gap at the $\Gamma$ point when spin$–$orbit coupling (SOC) is not considered [Fig. \textcolor{blue}{1(b)}]. The valence band around the $\Gamma$ point near the Fermi level is characterized as Mexican hat. With SOC, the Mexican hat-like valence band turns out to be cone-like, as shown in Fig. \textcolor{blue}{1(c)}. Thus the material is a semiconductor with a direct band gap. Such a unique cone-like band structure with very small effective mass of carriers is rather rare for 2D materials obtained from the layered bulk parents with the space group of $R\overline{3}m$.

\vspace{-2mm}
\subsection{Effect of spin-orbit coupling on energy band dispersion}
\vspace{-2mm}

In this section, we demonstrate why and how SOC can dramatically transform the band dispersion. It is necessary to capture the essential physics of the nontrivial band structure at $\Gamma$ point near the Fermi level by atomic orbital energy level diagram in conjunction with symmetry group. Atomic-projected band structure illustrates that the valence band maximum (VBM) and the conduction band minimum (CBM) are dominantly contributed by Te layer and Bi/Se double bilayers, respectively, as shown in Appendix \textcolor{blue}{B}. This is also supported by the partial charge density in Appendix \textcolor{blue}{B}, which is well consistent with our previous study \cite{wang2020janus}. 

According to the symmetry analysis and calculation of irreducible representations \cite{GAO2021107760}, we can infer that the VBM is derived from Te $p_{x,y}$ orbitals, while the CBM originates mainly from Bi $p_{z}$ orbitals. In Fig. \textcolor{blue}{2(a)}, an atomic orbital energy level diagram is drawn on the basis of above dicussion. Due to the geometric feature, Bi/Se double bilayers are separated by Te layer, so that the Bi/Se energy levels are lifted up while that of Te is pushed down due to the repulsion from both sides (stage I). Bonding and anti-bonding states form because of the interactions between Se/Bi double bilayers and Te layer, and the antibonding orbitals are higher in energy than the bonding orbitals. In the presence of inversion symmetry, bonding and antibonding orbitals have certain parities. Considering the above analysis, $\vert BiSe_{x,y,z}^+ \rangle$ and $\vert Te_{x,y,z}^- \rangle$ can be addressed to be near the Fermi level and we focus on these states for further discussion (stage II). QL Bi$_{2}$Se$_{2}$Te belongs to the $p\bar{3}m1$ layer group, thus the transformation properties of $z$ component are different from $x$,$y$ components of arbitrary vector under the $D_{3d}$ symmetry operation. Under the action of the crystal field, the $p$ components of both $\vert Te^- \rangle$ and $\vert BiSe^+ \rangle$ states split into $p_{z}$ orbital and doubly degenerate $p_{x,y}$ orbitals: $\vert Te^-,p_{x,y} \rangle$ states are higher in energy than $\vert Te^-,p_{z} \rangle$ state, while the energy of $\vert BiSe^+,p_{x,y} \rangle$ states is lower than that of $\vert BiSe^-,p_{z} \rangle$ state. In this case, the VBM is thus mainly composed of $\vert Te^-,p_{x,y} \rangle$ and the CBM is dominated by the $\vert BiSe^+,p_{z} \rangle$ (stage III). 

The SOC Hamiltonian is given by $\hat{H}=\lambda\bm{S}\cdot\bm{L}$ that couples the orbital angular momentum to spin, with $\lambda$ being the coupling strength. Experiencing the SOC, $\vert Te^-,p_{x,y} \rangle$ states split into two doubly degenerate states, namely, $\vert Te^-,\pm\frac{3}{2} \rangle$ of higher energy and $\vert Te^-,\pm\frac{1}{2} \rangle$ with lower energy (stage IV). Variation of the dispersion relationship around the $\Gamma$ point as a function of SOC strength intuitively shows the significant impact of SOC (see Appendix \textcolor{blue}{B}). Therefore, it is the SOC of Te atom that dominates the change in band dispersion, leading to the indirect$–$direct transition of the band gap.

\begin{figure}
	\centering
	\includegraphics{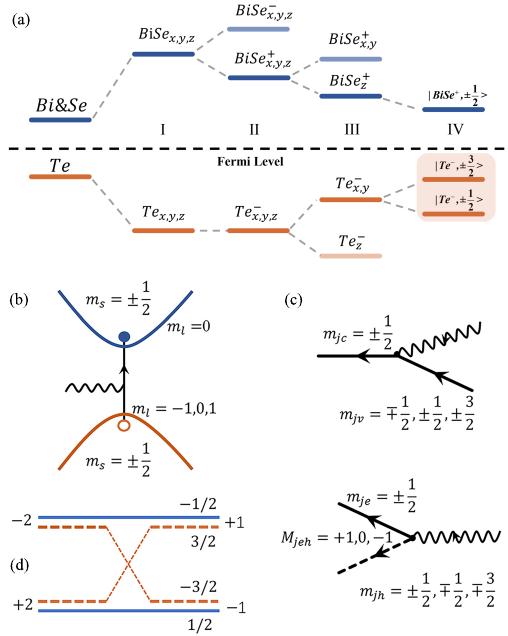}
	\caption{(a) Atomic orbital energy level diagram. (b) Spin-conserved photon absorption. (c) One photon combining one valence electron gives one conduction electron (upper panel), this can be described as an electron–hole pair creation process (lower panel). (d) Carrier exchange between two excitons, which converts bright excitons into dark excitons.}
	\label{FIG.2.}
\end{figure}

\vspace{-2mm}
\subsection{Optical properties and excitonic effect}
\vspace{-2mm}
QL Bi$_{2}$Se$_{2}$Te exhibits desirable ground-state electronic properties. However, the transitions between valence and conduction band edges may be parity-forbidden due to the inversion symmetry of the material, which could seriously deteriorate the optical absorption. Therefore, whether the transition is allowed by symmetry or not is a key factor in judging the absorption. For QL Bi$_{2}$Se$_{2}$Te, opposite parity between the VBM and CBM states allows the transitions. Transition dipole moment $\langle\varphi_{h}\vert\hat{\bm{\mu}}\vert\varphi_{e}\rangle$ denotes a transition between the initial state $\vert\varphi_{h}\rangle$ and the final state  $\vert\varphi_{e}\rangle$, with $\hat{\bm{\mu}}$ being the electric dipole operator \cite{meng2017parity,luo2018efficient}. The transition dipole moment determines how the material will interact with the electromagnetic wave of a given polarization, and the square of the magnitude gives the transition probabilities between two states. For QL Bi$_{2}$Se$_{2}$Te, the transition dipole moment amplitude at $\Gamma$ point is as large as 2055 Debye$^2$, implying significant optical absorption, as shown in the inset to Fig. \textcolor{blue}{3(a)}.

\begin{figure}
	\centering
	\includegraphics{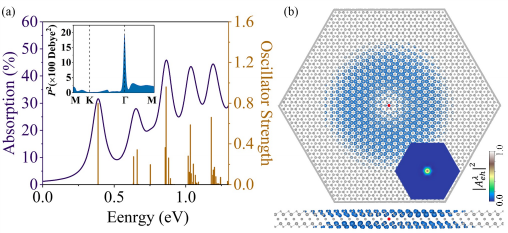}
	\caption{(a) Optical absorption of QL  Bi$_{2}$Se$_{2}$Te, with the oscillator strength depicted by brown strips. Inset presents the transition matrix elements. (b) Real-space exciton wave function distribution plotted over a 51$\times$51$\times$1 supercell, the hole is fixed on Te layer (red dot). Inset stands for the reciprocal-space distribution of the lowest-energy exciton at each reciprocal lattice point.}
	\label{FIG.3.}
\end{figure}

Considering the many-body effects and the full relativistic SOC, optical absorption spectrum of QL Bi$_{2}$Se$_{2}$Te is obtained by solving the Bethe$–$Salpeter equation (BSE) \cite{marini2009yambo}. In Fig. 3(a), the lowest-energy peak around 0.39 eV corresponds to four bound excitons, where the doubly degenerated dark excitons are 0.05 meV lower in energy than the doubly degenerated bright excitons. The exciton binding energy as high as 0.59 eV due to single-particle band gap is 0.98 eV in G$_0$W$_0$@PBE level. Such an excitation process can be described by the angular momentum selection rule for the absorption of one photon, giving rise to the valence-to-conduction band transition. Because of spin conservation, electron keeps its spin $m_{s}=\pm\frac{1}{2}$ after excitation, while its orbital angular momentum changes from ($l$=1, $m_{l}=0,\ \pm$1) to ($l$=0, $m_{l}$=0), see Fig. 2(b). It readily shows that the total angular momentum is $M_{jeh}=m_{je}+m_{jh}=(0,\ \pm1)$ for the electron–hole pair generated by absorbing one photon ($\pi,\ \sigma_{\pm}$), Fig. \textcolor{blue}{2(c)}. However, Pauli exclusion causes carrier exchange because of the composite nature of excitons \cite{pikus1971exchange,combescot2007bose}, thus the bright excitons with spin $\pm1$ can be transformed into dark excitons with spin $\pm2$ through carrier exchange, Fig. \textcolor{blue}{2(d)}.

To verify the composition and symmetry of these exciton states, exciton wave function in $k$-space is obtained. Our results indicate that the weight is overwhelmingly localized in vicinity of the $\Gamma$ point, indicating a direct transition between the VBM and CBM states, as shown in the inset to Fig. \textcolor{blue}{3(b)}. In addition, exciton wave function distribution in real space attained by diagonalizing the BS two-particle Hamiltonian gives more information regarding the localization and symmetry of the bound exciton, see Fig. \textcolor{blue}{3(b)}. Particularly, the exciton distribution probability confirms the spatial separation of quasi-electrons and quasi-holes, which are confined into Bi/Se double bilayers and Te layer, respectively. This is in agreement with the results of ground state. It can be found that the lowest-energy exciton spreads over almost eight unit cells, which is an important result for our following discussion. We estimated the mean square radius ($r_{RMS}$) of the exciton as the in-plane gyration radius, approximately 3.2 nm (see Appendix \textcolor{blue}{C}); this is larger than the lowest-energy excitons in TMDCs monolayers \cite{zipfel2018spatial,goryca2019revealing,dong2021direct}.

\vspace{-2mm}
\subsection{Excitonic phase transition.} 
\vspace{-2mm}

QL Bi$_{2}$Se$_{2}$Te thus corresponds to a quite small critical density of $n_{c}\approx\frac{1}{\pi r_{RMS}^2}\sim3.11\times{10}^{12}$ cm$^{-2}$ due to the large exciton radius. The critical density is greatly smaller than those of all reported vdW heterostructures \cite{fogler2014high,gupta2020heterobilayers,ulman2021organic}. In general, the critical density is set as the highest density limit for the boson condensate state. According to the criterion of $n_{M}r_{RMS}^{2}\sim0.09$, the Mott critical density $n_{M}$ for QL Bi$_{2}$Se$_{2}$Te can be attained as 0.88$\times{10}^{12}$ cm$^{-2}$, which is a relatively small value. A natural coniecture is that with such a dilute density, the BEC-like condensate phase of the excitons can only occur below a very small critical temperature \cite{fogler2014high,hu2021polariton}. While, it is not the case. It is necessary to clarify that dilute density depicts the strength scaling relationship between inter-exciton interaction and other energy scales. With dilute density condition is satisfied ($nr_{ex}^2\ll1$), the average distance between excitons is much larger than their size and the range of interaction. As a result, the interactions between excitons are weaker than other energy scales. In addition to the Mott critical density, however, the exciton effective mass also plays a decisive role in determining the characteristic temperature for excitons to become degenerate. The standard expression correlating the exciton effective mass and the characteristic temperature is \cite{fogler2014high,ulman2021organic,guo2022tuning} 
\begin{equation}\label{eq1}
	k_{B}T_{d}=\frac{{2\pi\hbar}^{2}}{M} n_{M},
\end{equation}
where $k_{B}$ is the Boltzmann constant, $T_{d}$ is the degeneracy temperature, $\hbar$ is the reduced Planck’s constant, and $M=m_{e}+m_{h}$ is the exciton effective mass. The cone-like band dispersion of QL Bi$_{2}$Se$_{2}$Te has already indicated the significantly small effective masses for both electrons and holes as well as small exciton effective mass at the $\Gamma$. Considering the quantum confinement and the reduced dielectric screening in low-dimensional materials, GW correction (strongly interacting electrons are described by weakly interacting quasi-particles) is applied to describing the band dispersion of QL Bi$_{2}$Se$_{2}$Te. With GW approximation included, the effective mass for quasi-electron is $m_{e}=0.09m_{0}$ and for quasi-hole $m_{h}=0.10m_{0}$, giving rise to an exciton effective mass ($M=0.19m_{0}$) of QL Bi$_{2}$Se$_{2}$Te which is much smaller than those of TMDCs monolayers \cite{berman2016high} [also see Appendix \textcolor{blue}{C}]. Thus, $T_{d}^{max}$ as high as 257 K can be attained for QL Bi$_{2}$Se$_{2}$Te.

Having determined the Mott critical density of 0.88$\times{10}^{12}$ cm$^{-2}$ and the Mott critical temperature of 257 K, a phase diagram for the electron$–$hole neutral QL Bi$_{2}$Se$_{2}$Te is constructed to visually illustrate the phase transition in the low exciton density regime [see Fig. \textcolor{blue}{4(a)}]. In the phase diagram, the oblique line across ($T_{d}^{max}$, $n_{M}$) borders the degenerate phase from the classical phase, that is as true for exciton gas as it is true for electron$–$hole gas; and the vertical line separates the exciton gases from the electron$–$hole gases at the critical Mott density $n_{M}$. At lower exciton density, excitonic BEC phase corresponds to strong coupling due to the reduced exciton screening. In contrast to BEC, Berezinskii$–$Kosterlitz$–$Thouless (BKT) phase has local phase coherence and can be characterized as superfluidity \cite{dong2021direct}. The phase transition temperature for BKT superfluid state is obtained from a universal, widely-accepted relation \cite{kosterlitz1973ordering, nelson1977universal} 
\begin{equation}\label{eq2}
	k_{B}T_{BKT}=\frac{{\pi\hbar}^{2}}{2M} n_{M}.
\end{equation}
The upper limit for $T_{BKT}$ at Mott critical density is estimated to 64.25 K for QL Bi$_{2}$Se$_{2}$Te. This is an exciting result: it is superior to all the systems experimentally and theoretically reported \cite{fogler2014high,gupta2020heterobilayers,wang2019evidence,ulman2021organic}. At low temperature, the carrier screening increases as the exciton density increases, leading to the phase transition from BKT to Bardeen–Cooper–Schrieffer (BCS) phase of exciton \cite{hu2021polariton,nozieres1985bose}. BCS phase corresponds to the weak coupling of composite Bosons, and holds when $n_{M}\textless n_{BCS}\le n_C$ with $n_{BCS}$ being the corresponding carrier or pair density. In this sense, BEC and BCS phases are two limits of the Boson condensate state. Additionally, the Mott transition can also occur above the Saha temperature
\begin{equation}\label{eq3}
	k_{B}T_{s}\sim\left(\frac{\pi{\hbar}^2}{M}n_{x}\right)e^{\frac{E_{b}}{k_{B}T_{s}}},
\end{equation}
where transition from classical exciton gas to classical electron–hole Fermi gas occurs (Not shown in Fig. \textcolor{blue}{4(a)} due to its high critical temperature ($>$ 400 K) in the low density region). 

According to the extended Beliaev theory \cite{Utesov053617}, remarkable exciton–exciton repulsive interactions can be anticipated for excitons with small mass. However, in addition to the effective mass, the inter-exciton interaction in 2D systems is also related to the density. As shown in Appendix \textcolor{blue}{D}, a repulsive interaction $V_{0}$ $=$ 3.02 $eV\cdot nm^{2}$ is obtained for QL Bi$_{2}$Se$_{2}$Te at the Mott critical density $n_{M}$, which is slightly larger than that of the MoSe$_{2}$–WSe$_{2}$ heterostructure \cite{Erkensten045426}.
 
 Moreover, as shown in Appendix \textcolor{blue}{D}, the critical temperatures of QL Bi$_{2}$Se$_{2}$Te are $T_{C}=263$ $K$ and $T_{BKT}=69.90$ $K$ for BEC and superfluid critical temperature, respectively, with inclusion of the excitonic interaction. They are slightly higher than that obtained from the standard formula. Therefore, we can conclude that the standard formula stated above can also describe the excitonic behavior well. Such agreement arises from the weak interaction between the excitons. 

\vspace{-2mm}
\subsection{Excitonic lifetime}
\vspace{-2mm}

In practice, only quasi-equilibrium state is possible because of exciton recombination, thus the limitation posed by a finite exciton lifetime is another ingredient in the discussion of excitonic condensation. We used the method proposed by Louie and Palummo et. al, which employs the Fermi’s Golden rule to derive the radiative decay rate $\gamma_{S}\left(\bm{Q}\right)$ of the exciton in state S with a wave vector of $\bm{Q}$ \cite{spataru2005theory,palummo2015exciton}. As the inverse of the decay rate, exciton radiative lifetime for $\bm{Q}=0$ can be written as
\begin{equation}\label{eq4}
	\tau_{S}\left(0\right)=\frac{\hbar^{2}cA_{uc}}{8\pi e^{2}E_{S}\left(0\right)\mu_{S}^2},
\end{equation}
where $c$, $A_{uc}$, and $E_{S}\left(0\right)$ represent the speed of light, area of the unit cell, and the excitation energy, respectively; $\mu_{S}^2$ is the square modulus of the exciton transition dipole divided by the number of $k$ points. Results indicate that the lowest-energy exciton ($X_{0}$) of QL Bi$_{2}$Se$_{2}$Te has a rather long radiative lifetime of $\tau_{0}\sim1.47$ ns at zero temperature. At a finite temperature, the average radiative lifetime in state $S$ is given as
\begin{equation}\label{eq5}
	\tau_{S}\left(T\right)=\tau_{S}\left(0\right)\cdot\frac{3}{4}\left(\frac{2Mc^{2}}{E_{S}^2\left(0\right)}\right)k_{B}T.
\end{equation}
In case of QL Bi$_{2}$Se$_{2}$Te, the radiative lifetime $\tau_{S}\left(257\right)$ for $X_{0}$ is approximately 31.24 $\mu s$. The long lifetime of exciton $X_{0}$ can be attributed to its characteristic spatial separation. It should be emphasized that only a desirable exciton radiative lifetime in a broad temperature range is encouragingly to unconditionally speak about the BEC/BKT phase.

\begin{figure}[H]
	\centering
	\includegraphics{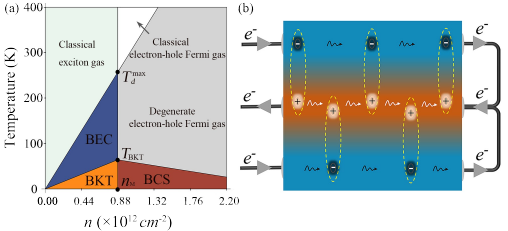}
	\caption{(a) Phase diagram for excitonic condensation of the lowest-energy exciton. $n$ is the excitonic density, $T$ is the temperature, and $n_{M}$ is the Mott critical density. Strictly speaking, the BEC region corresponds to excitonic quasi-condensation instead of condensation. This is due to that unambiguous distinguishing between them is hard in experiments, whereby we mark the dark blue area as BEC. (b) Schematic of exciton superfluid transistor.}
	\label{FIG.4.}
\end{figure}

\vspace{-2mm}
\section{Discussion}
\vspace{-2mm}

Apart from the phases discussed above, exciton gas can also transfer to electron–hole liquid (EHL) state, rather than the excitonic condensation. If quasi-electrons and quasi-holes are spatially separated, however, excitons behave as oriented electric dipoles with repulsive interaction, which is crucial to prevent the occurrence of EHL state and promote the formation of BKT/BCS phase at low densities \cite{butov2007cold}. The EHL state can be more stable at higher densities \cite{wu2015theory}.

Generally, it is difficult to determine intuitively whether the electron–hole pairs in a semiconductor form unbound fermionic particles or thermalized excitons when they are excited. The determination of the dominating phase is associated with many factors, such as temperature, dielectric environment as well as excitation density, \textit{etc.} \cite{Katsch257402,Steinhoff2017}. In addition, unbound fermionic particles may further hinder the formation of excitons due to the Pauli blocking effects. However, the research of Steinhoff \textit{et al.} pointed out that the degree of the ionization of the system depends greatly on the excitation density, and excitons play a dominating role on the low side of Mott density  \cite{Steinhoff2017}. In addition, changing the dielectric environment of the system can eliminate the influence of ionization and ensure that the system obeys the Bose–Einstein statistics. We found that the Mott density obtained by Fogler \textit{et al.} is in good agreement with that of Steinhoff \textit{et al.}, with the former slightly smaller than the latter while in the same order of magnitude \cite{fogler2014high, Steinhoff2017}. Thus, we can infer that it is a reliable way to determine the Mott density of QL Bi$_{2}$Se$_{2}$Te base on the method of Fogler et al, and this also means that the estimation of the maximum BEC and superfluidity critical temperature does make sense. Moreover, Steinhoff \textit{et al.} demonstrated that full excitonic ionization of TMDCs with SiO$_{2}$ substrate requires temperature up to 300 K in electron–hole excitation densities close to zero. Such a temperature is much higher than the temperature at which the low excitation density in TMDC systems without substrate can be theoretically predicted to achieve BEC and superfluidity. Therefore, it can be expected that BEC and superfluidity can well exist above the room temperature in suitable dielectric environments. Of course, this way also applies to QL QL Bi$_{2}$Se$_{2}$Te system.

In addition, in case the interlayer distance between electron and hole is smaller than the exciton Bohr radius, biexcitons may probably form and a solid-state setting can be used to achieve resonantly paired bosonic superfluid \cite{Meyertholen235307,Andreev140501}. In QL Bi$_{2}$Se$_{2}$Te, however, polarized light can be used as an efficient tuning knob for the formation of biexcitons. In particular, linearly polarized supports the formation of biexcitons, rather than the $\sigma_{+}$ or $\sigma_{-}$ light \cite{Katsch257402}. Small exciton mass can also lead to smaller biexciton binding energy than that of exciton liquid, where biexciton is not beneficial \cite{lozovik1997phase}. A great deal of effort has been also devoted to realizing dipolar polariton-based quantum devices, for which the tunable interaction is particularly important \cite{Andreev125129}. On the other hand, because of the low exciton effective mass, the width of the shape resonance in the exciton–exciton interaction is expected to be large, and, therefore, resonant pairing of such excitons would be similar to the resonant pairing of dipolar polaritons \cite{Andreev125129}. Strong correlations and squeezing of emitted photons could potentially be achievable within the free-standing setting constructed by QL  Bi$_{2}$Se$_{2}$Te.

\vspace{-2mm}
\section{Conclusion}
\vspace{-2mm}

In summary, we show that QL Bi$_{2}$Se$_{2}$Te is an unprecedented choice to study high-temperature excitonic condensation because of the unique features of its indirect excitons. In terms of applications, bound excitons travelling in material as a superfluid with zero viscosity offer a great promise to achieve room-temperature dissipationless optoelectronic devices, see Fig. \textcolor{blue}{4(b)} of the schematic for the exciton transitor based on QL Bi$_{2}$Se$_{2}$Te. In particular, electrons are distributed in the top and bottom Bi/Se bilayers, while holes are confined within the middle Te layer; the electrons and holes bind together forming excitons. When the electrons are accelerated by an applied voltage, the partnering holes can be dragged by its electrons. QL Bi$_{2}$Se$_{2}$Te thus provides a potential platform where excitons can be formed and condensed into superfluidity to flow without resistance, carrying electric current in opposite direction without dissipation.

\textit{Note added.} During the submission process we became aware of similar work and materials on related topics in Refs. \cite{afm202303779, PENG2024386}. 
\vspace{0.2cm}
\begin{center}
	\noindent{\bf ACKNOWLEDGMENTS}
\end{center}
\vspace{0.2cm}
This work is supported by the National Natural Science Foundation of China (52272223 and 12074217). Y. W. and Y. S. A. are also supported by the Singapore Ministry of Education (MOE) Academic Research Fund (AcRF) Tier 2 Grant under the award number MOE-T2EP50221-0019. The computational work for this article was performed partially on resources of the National Supercomputing Centre, Singapore (https://www.nscc.sg).

\vspace{0.2cm}
\begin{center}
	\noindent{\bf APPENDIX A: EXFOLIATION ENERGY AND  STRUCTURE STABILITY }
\end{center}

\begin{figure*}
	\centering
	\includegraphics{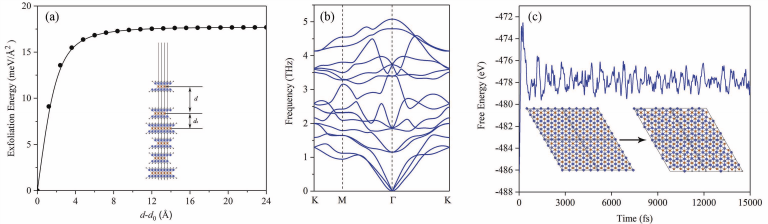}
	\caption{(a) Exfoliation energy for QL Bi$_2$Se$_2$Te, the inset shows the side view of bulk Bi$_2$Se$_2$Te and the interlayer distance. (b) Phonon band dispersion of QL Bi$_2$Se$_2$Te over the first Brillouin zone. (c) Free energy variation of QL Bi$_2$Se$_2$Te at 500 K during the AIMD simulations, insets show the equilibrium structures at 500 K.}
	\label{FIG.5.}
\end{figure*}

To obtain accurate estamation of exfoliation energy, DFT–D3 corrections within the PBE functional were taken into account \cite{graziano2012improved}. In the inset of Fig.~\textcolor{blue}{5}, the procedure of mimicking experimentally mechanical exfoliation is illustrated. It can be found from Fig.~\textcolor{blue}{5(a)} that the exfoliation energy for QL Bi$_{2}$Se$_{2}$Te was calculated to be 17.68 meV/\AA$^{2}$, which is comparable to that of graphene (12 meV/\AA$^{2}$) and MoS$_{2}$ (26 meV/\AA$^{2}$) \cite{bjorkman2012van,mounet2018two}. It is an indication of high feasibility to experimentally exfoliate QL Bi$_{2}$Se$_{2}$Te from the layered bulk material.

In order to check the dynamic stability of QL Bi$_{2}$Se$_{2}$Te, we calculated the phonon spectrum based on a 5$\times$5$\times$1 supercell in framework of density functional perturbation theory by using the Phonopy code \cite{togo2015first}. Fig.~\textcolor{blue}{5(b)} shows the positive frequencies for all phonon branches, a sign confirming the dynamic stability of QL Bi$_{2}$Se$_{2}$Te. In addition, ab initio molecular dynamics (AIMD) simulations were performed based on a 5$\times$5$\times$1 supercell at 500 K for 15 ps with a time step of 3 fs. As shown in Fig.~\textcolor{blue}{5(c)}, the free energy fluctuates slightly within a small range and the configuration can be well kept when annealing at 500 K for 15 ps, which are strong evidences for the thermal stability of QL Bi$_{2}$Se$_{2}$Te.

\vspace{0.2cm}
\begin{center}
	\noindent{\bf APPENDIX B: ELECTRONIC PROPERTY }
\end{center}

Atomic-projected band structure illustrates that the valence band maximum (VBM) and the conduction band minimum (CBM) are dominantly contributed by Te layer and Bi/Se double bilayers, respectively, Fig.~\textcolor{blue}{6(a)}.  And it is supported by the partial charge density as shown in inset of Fig.~\textcolor{blue}{6(a)}. In Fig.~\textcolor{blue}{6(b)}, variation of the band structure around the $\Gamma$ point is shown as a function of SOC strength from formally 0\% (without SOC) to 100\% (with SOC), which intuitively presents the significant impact of SOC. Therefore, it is the SOC of the Te atom that dominates the change in the band dispersion, leading to the indirect–direct transition of the band gap. In addition,  Fig.~\textcolor{blue}{7} shows $C_{3}$ symmetric energy dispersion relationship of QL Bi$_2$Se$_2$Te over the first Brillouin zone without and with SOC included.

\begin{figure}
	\centering
	\includegraphics{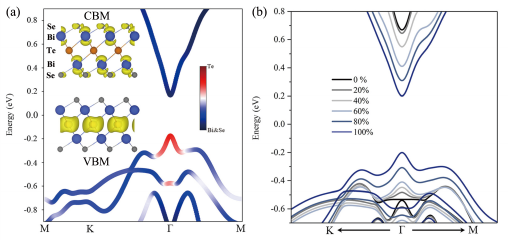}
	\caption{(a) Atomic-projected band structure of QL Bi$_{2}$Se$_{2}$Te, inset is the partial charge densities, isosurface value is set to 0.003 eV/\AA$^{3}$. (b) Band structure evolution as a function of SOC strength.}
	\label{FIG.6.}
\end{figure}

\begin{figure}
	\centering
	\includegraphics{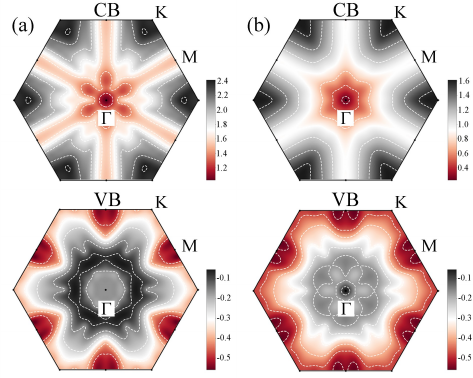}
	\caption{2D band structure of QL Bi$_{2}$Se$_{2}$Te in the first Brillouin zone. (a) Band structure without considering SOC. (b) Band structure with SOC is included}
	\label{FIG.7.}
\end{figure}

\vspace{0.2cm}
\begin{center}
	\noindent{\bf APPENDIX C: EXCITON ROOT MEAN SQUARE RADIUS AND EFFECTIVE MASS}
\end{center}

The mean square radius of the exciton is defined by the radial exciton wave function $\psi_{n}(\bm{r})$ and the electron–hole separation $\bm{r}-\bm{r}_0$, using the formula \cite{Zipfel075438}
\begin{equation}
	\langle r^{2}\rangle=\langle\psi_{n}\vert\left(\bm{r}-\bm{r}_{0}\right)^{2}\vert\psi_{n}\rangle=\frac{\int_{0}^{\infty}{\left(\bm{r}-\bm{r}_{0}\right)^{2}\vert\psi_{n}\left(\bm{r}\right)\vert^{2}\bm{r} d\bm{r}}} {\int_{0}^{\infty}\vert\psi_{n}\left(\bm{r}\right)\vert^{2}\bm{r}d\bm{r}},
\end{equation}
with $r_{0}$ being the position of hole. The root mean square (RMS) radius $r_{RMS}=\sqrt{\langle r_{n}^{2}\rangle}$ characterizes the spatial extent of the exciton.
In general, the root mean square (gyration radius) $r_{RMS}$ is given by $ r_{RMS}^2=\left\langle\varphi_{n}\middle|{\hat{r}}^2\middle|\varphi_{n}\right\rangle$, while the Bohr radius of exciton is calculated as $ r_{ex}=\left\langle\varphi_{n}\middle|\hat{r}\middle|\varphi_{n}\right\rangle$. Here, $\varphi_{n}(r)$ is the radial exciton wave function with $r$ being the electron$–$hole separation. For 2D systems, if making the analogy with a 2D hydrogen atom \cite{yang1186} with principle quantum number $n$, we can obtain
\begin{equation}
	\left\langle n\middle|{\hat{r}}^2\middle|n\right\rangle =\frac{1}{8}\left(2n-1\right)\left[n\left(10n^2-15n+11\right)-3\right]a_{0}^2
\end{equation}
\begin{equation}
	\left\langle n\middle|\hat{r}\middle| n\right\rangle=\frac{1}{2}[3n(n-1)+1]a_{0}
\end{equation} 
As the principal quantum number $n = 1$, for the low-energy states we have $\frac{r_{RMS}}{r_{ex}}=\frac{\sqrt{3/8}}{1/2} \sim 1.225$.
The critical density $n_{M}$, below which excitons behave as weakly interacting Bosons, is the key for the excitonic BEC phase transition. In our work, we considered the condition of the dilute limit reported in the empirical formulas in 2D cases. If the ratio between the in-plane gyration radius $r_{RMS}$ and the mean inter-exciton distance reaches a critical value of about 0.3 ($\frac{r_{RMS}}{1/\sqrt{n_M}}\approx0.3$) \cite{de206401, maezono216407, fogler2014high}, exciton dissociation will occur. Accordingly, the corresponding critical density $n_{M}$ should satisfy $n_{M}r_{ex}^2\approx0.06$. This is in excellent consistent with Hohenberg’s theory \cite{fisher4936}.

The effective mass is nearly isotropic for QL Bi$_{2}$Se$_{2}$Te around the point $\Gamma$, due to the $C_{3}$ symmetry as shown in Fig.\textcolor{blue}{7b}. As a result, the band dispersion relation can be locally approximated as  $E(k)=E_{0}+\hbar^2 k^2 /2m^*$. In accordance to $m^*=\hbar^{2}\left[\partial^{2}E(k)/{\partial k}^2\right]^{-1}$, electron and hole effective masses of QL Bi$_{2}$Se$_{2}$Te around $\Gamma$ point can be calculated by the band dispersion. Once electron/hole effective masses $m_{zig}$ and $m_{arm}$ was obtained along zigzag and armchair directions, respectively, then the carrier effective mass can also be obtained by $m_{eff}=(m_{zig}+m_{arm})/2$. In our calculations, SOC was considered. In TABLE I, results are summarized. 

\begin{table}[H]
	\centering
	\caption{Carrier effective mass for QL Bi$_{2}$Se$_{2}$Te, the unit is electron mass $m_{0}$.}
	\label{1}
	\begin{tabular}{c|c|c|c}
		\hline
		\hline
		carrier &~~~$m_{zig}$~~~ &~$m_{arm}$~ &~$m$ \\
		\hline                                       
		hole    &0.10  &0.10  &0.10 \\
		electron   &0.09  &0.09  &0.09 \\	
		\hline\hline
	\end{tabular}
\end{table}

Based on these expressions mentioned above, results of carrier effective mass have been proved to be exactly correct in many previous publications. We also verified that it is correct for cases of 2H-TMDCs, which show band dispersion relations similar to the massive Dirac cone around the band edge: 0.3$–$0.5 $m_{0}$, in good agreement with previous results \cite{cheng8b07871}. Therefore, we confirm that the parabolic approximation can be applied to fit the band edge accurately for the QL Bi$_{2}$Se$_{2}$Te also with a massive Dirac cone.

\vspace{0.2cm}
\begin{center}
	\noindent{\bf APPENDIX D: PHASE TRANSITION TEMPERATURE OF WEAKLY INTERACIING EXCITON }
\end{center}

Next, we will give a effective model to show that, in two-dimensional QL Bi$_{2}$Se$_{2}$Te, the weakly interacting bosons with small exciton effective mass can realize the excitonic BEC/superfluid state in high critical temperature.

In a dilute limit, excitons are featured by weak interactions, which affect the excitonic properties of QL Bi$_{2}$Se$_{2}$Te in a dramatic way. In QL Bi$_{2}$Se$_{2}$Te without introducing external field, without loss of generality, we assume that excitons are uniformly distributed in a 2D container with side length $L$, which is required to satisfy the periodic boundary conditions. Then, the Hamiltonian of the system can be written as	
	\begin{equation}
		\hat{H}=\sum \frac{p^{2}}{2M}\hat{a}^{\dagger}_{\bm{p}}\hat{a}_{\bm{p}}+\frac{1}{2L^{2}}\sum{V_{\bm q}\hat{a}^{\dagger}_{\bm{p_{1}}+\bm{q}} \hat{a}^{\dagger}_{\bm{p_{2}}-\bm{q}} \hat{a}_{\bm{p_{1}}} \hat{a}_{\bm{p_{2}}}},
	\end{equation}
	where $\hat{a}_{\bm{p}}$/$\hat{a}^{\dagger}_{\bm{p}}$ is the operator annihilating/creating a particle in the single-particle state with momentum $\bm{p}$, and $\bm{p}$ satisfies the usual periodic boundary conditions $M$ is the exciton effective mass. In the expression of the Hamiltonian,
	\begin{equation}
		V_{\bm{q}}=\int V(\bm{r})e^{-i\bm{q}\cdot \frac{\bm{r}}{\hbar}}d\bm{r},
	\end{equation}
	with $V(\bm{r})$ being the two-body potential. In regard to a real system, the inter-particle potential is always difficult to obtain by solving the Schrödinger equation at the microscopic level. In the dilute limit, however, the exact form of the two-body potential is not important for describing the macroscopic properties of the system, as long as it gives the correct value of the s-wave scattering length.In order to attain the simplest many-body formalism, it is customary to replace the microscopic potential $V$ with an effective potential $V_{eff}$ \cite{becbook, bec7216}. Since only small momenta are involved in the solution of the many-body problem, it is allowed to consider only $\bm{q}=0$ in the Fourier transform of $V_{eff}$, as $ V_{0}=\int V_{eff}(\bm{r})d\bm{r}$. In this case, therefore, the Hamiltonian can be rewritten as
	\begin{equation}
		\hat{H}=\sum \frac{p^{2}}{2M}\hat{a}^{\dagger}_{\bm{p}}\hat{a}_{\bm{p}}+\frac{V_{0}}{2L^{2}}\sum\hat{a}^{\dagger}_{\bm{p_{1}}+\bm{q}} \hat{a}^{\dagger}_{\bm{p_{2}}-\bm{q}} \hat{a}_{\bm{p_{1}}} \hat{a}_{\bm{p_{2}}}.
	\end{equation}
	In consideration of only the low-energy excited states, operators $\hat{a}^{\dagger}_{0}$ and $\hat{a}_{0}$ can be replaced by a $c$-number $\sqrt{N}$ according to the Bogoliubov approximation, where $N$ stands for the total number of the particles in system. By retaining all the quadratic terms in the particle operators with $\bm{p}\neq 0$, the expansion of the Hamiltonian yields
	\begin{equation}
		\hat{H}=\sum \frac{p^{2}}{2M}\hat{a}^{\dagger}_{\bm{p}}\hat{a}_{\bm{p}}+\frac{N^{2}V_{0}}{2L^{2}}+\frac{NV_{0}}{2L^{2}}\sum_{p\neq0}(2\hat{a}^{\dagger}_{\bm{p}}\hat{a}_{\bm{p}}+\hat{a}^{\dagger}_{\bm{p}}\hat{a}^{\dagger}_{-\bm{p}}+\hat{a}_{\bm{p}}\hat{a}_{-\bm{p}}).
	\end{equation}
	In particular, Eq.~(12) can be diagonalized on the basis of the Bogoliubov transformation,
	\begin{equation}
		\hat{a}_{\bm{p}}=u_{\bm{p}}\hat{b}_{\bm{p}}+v_{-\bm{p}}\hat{b}^{\dagger}_{-\bm{p}},	
		\hat{a}^{\dagger}_{\bm{p}}=u_{p}\hat{b}^{\dagger}_{\bm{p}}+v_{-\bm{p}}\hat{b}_{-\bm{p}}.
	\end{equation}
	The new operators, $\hat{b}_{\bm{p}}$ and $\hat{b}^{\dagger}_{\bm{p}}$, are assumed to obey the same bosonic commutation relation as the real particle operators of $\hat{a}_{\bm{p}}$ and $\hat{b}^{\dagger}_{\bm{p}}$. The commutation relation imposes the constraint for the two parameters $u_{\bm{p}}$ and $v_{-\bm{p}}$ with $u^{2}_{\bm{p}}-v^{^2}_{-\bm{p}}=1$. By virtue of the transformation, the coefficients of the non-diagonal terms $\hat{a}^{\dagger}_{\bm{p}}\hat{a}^{\dagger}_{-\bm{p}}$ and $\hat{a}_{\bm{p}}\hat{a}_{-\bm{p}}$ in Eq.~(12) vanish. Thus, the Hamiltonian can finally be reduced to the diagonal form
	\begin{equation}
		\hat{H}=E_{0}\sum\epsilon(p)\hat{b}^{\dagger}_{\bm{p}}\hat{b}_{\bm{p}},
	\end{equation}
	where $E_{0}$ represents the ground state energy, and $\epsilon(p)$ symbolizes the dispersion of exciton as
	\begin{equation}
		\epsilon(p)=\bigg[\frac{V_{0}n}{M}p^{2}+\big(\frac{p^{2}}{2M}\big)^{2}\bigg]^{1/2}
	\end{equation}
	with $n$ being the exciton density. In the light of the conservation of particle number in the bosonic system, we further obtain
	\begin{equation}
		\frac{Mk_{B}T_{C}}{2\pi\hbar^{2}}\int_{0}^{+\infty}dx\frac{x}{(e^{x}-1)\sqrt{(\frac{V_{0}n}{k_{B}T_{C}})^{2}+x^{2}}} =n
	\end{equation}
	It can be found that $\frac{V_{0}n}{k_{B}T_{C}}\neq0$, and, therefore, the integral is integrable.

	By fitting the exciton dispersion of QL Bi$_{2}$Se$_{2}$Te based on Eq. (15), we can get $V_{0}=3.02$ $eV\cdot nm^{2}$ which is slightly larger than that of the  MoSe$_{2}$–WSe$_2$ \cite{Erkensten045426}. Further, the BEC critical temperature of $T_{C}=263$ $K$ can be attained at $n_{M}$ by self-consistently solving Eq. (16). This result is consistent very well with the result in the main text ($T_{d}^{max}=257$ $K$) via the simplified model in Eq.~(1).

Taking the long-wave limit into account, on the other hand, $\epsilon(\bm{p})$ can be reduced to $ \epsilon(p)=\sqrt{\frac{V_{0}n}{M}}p$. In accordance to Landau’s criterion of superfluidity, excitons in weakly interacting quantum Bose system indicate zero viscosity when $v\leq\epsilon(p)/\pi p$, and the highest superfluid critical temperature $T_{BKT}$ satisfies the following relation
	\begin{equation}
		\frac{(k_{B}T_{BKT})^{2}}{2\pi\hbar^{2}\xi^{2}}\int_{0}^{+\infty}\frac{xdx}{(e^{x}-1)} =n
	\end{equation} 
	where $\xi=\sqrt{\frac{V_{0}n}{\pi^{2}M}}$. After substituting $C=1.645$ with the integral, thus, we can rewrite the expression as
	\begin{equation}
		k_{B}T_{BKT}=\bigg(\frac{2\pi\hbar^{2}\xi^{2}n}{C}\bigg)^{1/2}
	\end{equation}
	Accordingly, the superfluid critical temperature turns out to be $T_{BKT}=69.90$ $K$ at $n_M$, which is consistent also very well with the result in the main text ($T_{BKT}=64.25$ $K$) via the simplified model in Eq.~(2).


%

\end{document}